# Status of technology of MRPC time of flight system

Yi Wang, Qiunan Zhang, Dong Han, Fuyue Wang, Yancheng Yu, Pengfei Lyu, Yuanjing Li

*Key Laboratory of Particle and Radiation Imaging, Department of Engineering Physics, Tsinghua University,Beijing 100084, China*

ABSTRACT: Time of flight system (TOF) based on MRPC technology is widely used in modern physics experiments, and it also plays an important role in particle identification. With the increase of accelerator energy and luminosity, TOF system is required to indentify definite particles precisely under high rate environment. The MRPC technology TOF system can be defined as three generations. The first generation TOF is based on float glass MRPC and its time resolution is around 80ps, but the rate is relatively low (typically lower than 100Hz/$cm^2$). The typical systems are TOF of RHIC-STAR, LHC-ALICE and BES III endcap. For the second generation TOF, its time resolution is in the same order with the first generation, but the rate capability is much higher. Its rate capability can reach 30kHz/$cm^2$. The typical experiment with this high rate TOF is FAIR-CBM. The biggest challenge is on the third generation TOF. For example, the momentum upper limit of K/PI separation is around 7GeV/c for JLab-SoLID TOF system under high particle rate as high as 20kHz/$cm^2$, the time requirement is around 20ps. The readout electronics of first two generations is based on time over threshold method and pulse shape sampling technology will be used in the third generation TOF. In the same time, the machine learning technology is also designed to analysis the time performance. In this paper, we will describe the evolution of MRPC TOF technology and key technology of each generation TOF.

**Contents**

**1. Introduction**

    Particle identification is very important in modern physics experiments. The method of time of flight (TOF) plays and important role in the identification of proton, kaon and pion. With the development of gas detectors, the Multi-gap Resistive Plate Chambers (MRPC) have been used to construct the TOF system such as RHIC-STAR, LHC-ALICE and et al [1,2]. The MRPC technology TOF system can be generalized as three generation. The first generation TOF is based on float glass MRPC and its time resolution is around 80ps, but the rate capability is relatively low (typically lower than 100Hz/cm$^2$). The typical systems are TOF of RHIC-STAR, LHC-ALICE and BES III endcap. For the second generation TOF, its time resolution is in the same order with the first generation, but the rate capability is much higher. The typical experiment with this high rate TOF is FAIR-CBM and its rate requirement can reach 30kHz/cm$^2$.The biggest challenge is on the third generation TOF. For example, the momentum upper limit of K/PI separation is around 7GeV/c for JLab-SoLID TOF system under high particle rate as high as 20kHz/cm$^2$, the time resolution requirement is around 20ps. The third generation can be defined as high rate and ultra high time resolution TOF system. The front end electonics is planed to be a fast amplifier and a charge digitizer, so the signal waveform is read out from the MRPC. We propose an end-to-end solution to obtain a more precise MRPC time with an artificial neural network(NN) to take full advantage of the waveform. We proved that this solution improves the time resolution greatly.This article is organized as follows: from Section 2 to 4, the three generation TOF is described. Section 5 summarizes the conclusions.

**2. The first generation TOF**

    The typical first generation TOF is for RHIC-STAR and LHC-ALICE. The typical feature is that the MRPC is assembled with float glass, the signal is amplified by fast differential amplifier and the time is obtained by TDC. The structure of MRPC module for STAR is shown in Fig.1. Each module consists of six read-out pads. The size of each pad is 3.1×6.0 cm$^2$. There is a 3 mm gap between each pad. The total active area of one module is 18.6×6 cm$^2$ and the working

electronegative gas consists of 95% F134a and 5% iso-butane. The electric field in the gas gap is larger than 100 kV/cm. The detector works in avalanche mode and has excellent time response. The total TOF consists of 3840 MRPC modules.

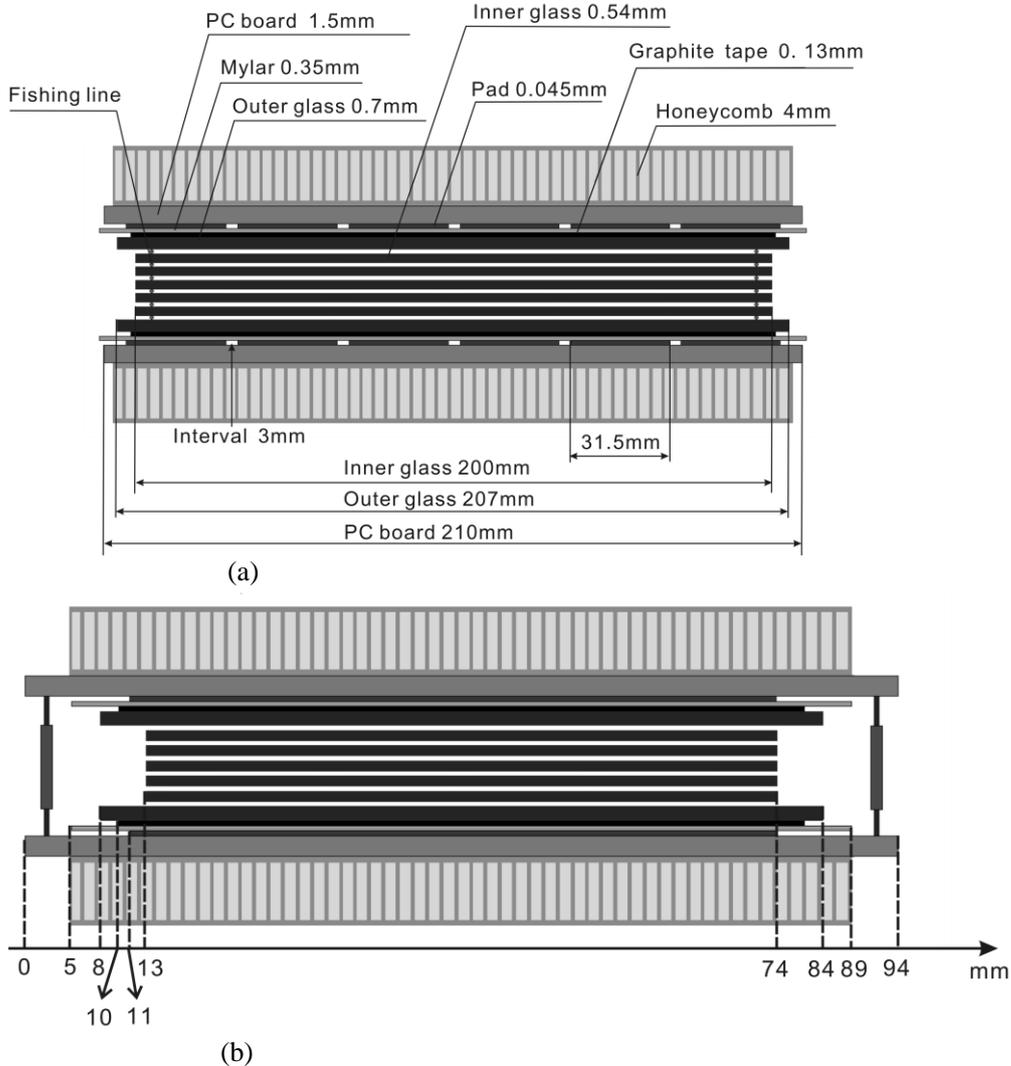

Fig. 1. Structure of the MRPC module for STAR-TOF.  (a) long side view  (b) short side view

Since a differential signal is obtained from the detector to give the best time resolution, it makes sense to use an amplifier/discriminator that has a differential input. An ASIC developed at CERN named the NINO are used [3]; The important aspects of this ASIC are: differential input; fast rise time (1 ns peaking time); low power consumption (45 mW/channel) and the input charge is encoded into the output width of the LVDS signal. The TDC used in STAR-TOF is also designed at CERN, the ASIC TDC is known as HPTDC [4]. This ASIC is fed a 40 MHz clock that is multiplied to a very high frequency using PLLs (phase locked loops) and delay locked loops. Its time resolution is around 25ps.

The typical particle identification of STAR-TOF is shown in Fig.2. It can be seen that the TOF detector has a high capability of separating p, K and π for momenta up to 3 GeV/c.

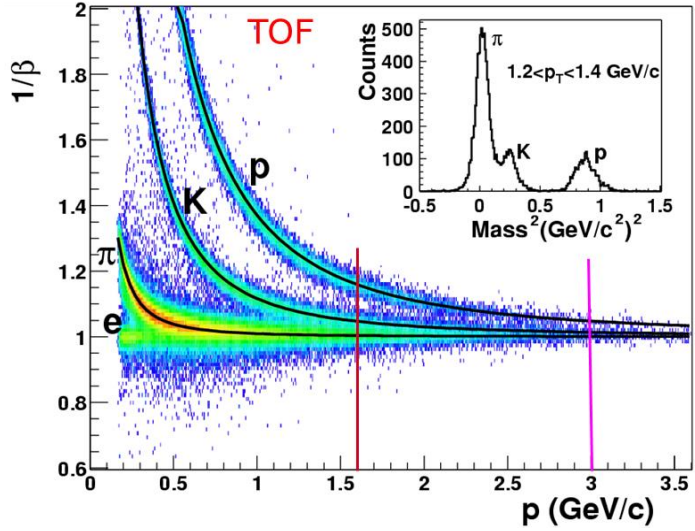

Fig.2 Particle identification provided by STAR-TOF

Plenty of physics results are obtained by the excellent PID of STAR-TOF. In 2011, the signal of antimatter Helium-4 [5] is observed in the Au+Au collision. Fig.3 shows isotope identification based on energy loss and mass calculated from momentum per charge and time of flight. The panels a and b of figure 3 shows the dE/dx versus the calculated mass. The panel a shows negatively charged particles, while the middle panel shows the positively charged particles. In both panels, the majority species are $^3$He and $^3\overline{\text{He}}$. The panel c shows the projection onto the mass axis of the top two panels. There is clear separation between $^3\overline{\text{He}}$ and $^4\overline{\text{He}}$ mass peaks.

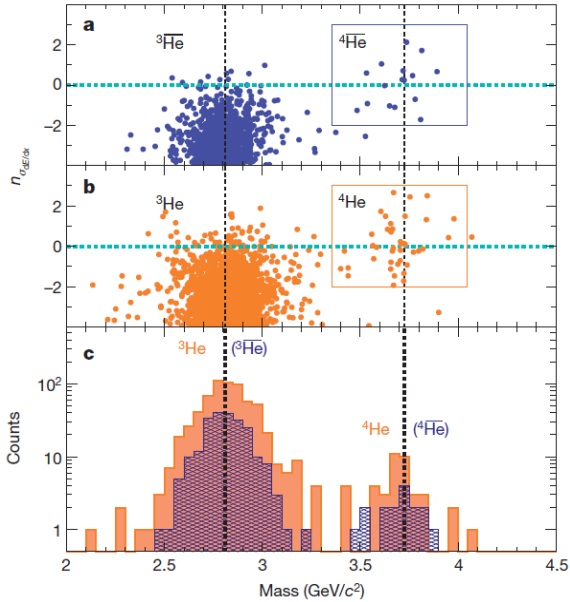

Fig.3 Isotope identification based on energy loss and mass calculated from momentum per charge and time of flight.

## 3. The second generation TOF

The typical second generation TOF is FAIR-CBM TOF. As we know, the Compressed Baryonic Matter (CBM) experiment at the Facility of Antiproton and Ion Research (FAIR) will be a high-rate fixed target experiment operated at ion beam intensity up to $10^9$ /s, which is sufficient to acquire data for rare probes such as charmed hadrons, multiple strange baryons, di-electrons and di-muons [6]. In a simulation of the Au + Au collision at such an interaction rate, the particle fluxes on the TOF wall can reach up to 30 kHz/cm$^2$. The CBM-TOF wall will be constructed with MRPC. This is a big challenge because the rate capability of MRPC assembled with float glass is lower than 500Hz/cm$^2$. The best way to improve the rate capability is to decrease the bulk resistivity of glass [7]. A kind of low resistive silicate glass (TUYK-LRG10) with bulk resistivity on the order of $10^{10}$ Ω cm was developed in Tsinghua University. This glass, characterized by an ohmic behavior and stability with transported charge, contains oxides of transition elements. It has a black color and is opaque to visible light. The main performance of the glass is shown in Table 1.

Table 1. Performance of the low-resistive glass

| Parameter | Typical value |
| --- | --- |
| Maximal dimension | 32 cm x 30 cm |
| Bulk resistivity | $10^{10}$ Ωcm |
| Standard thickness | 0.7, 1.1 mm |
| Thickness uniformity | 20 μm |
| Surface roughness | < 10 nm |
| Dielectric constant | 7.5 – 9.5 |
| DC measurement | Ohmic behavior stable up to 1 C/cm$^2$ |

Two MRPC prototypes based on such low-resistive glasses were produced and tested in ELBE beam test, Dresden-Rossendorf, Germany, to examine their performance under high rate. As shown in Fig 4, the efficiency is still higher than 90% and the time resolution is about 80ps, even though at 70 kHz/cm$^2$ rate.

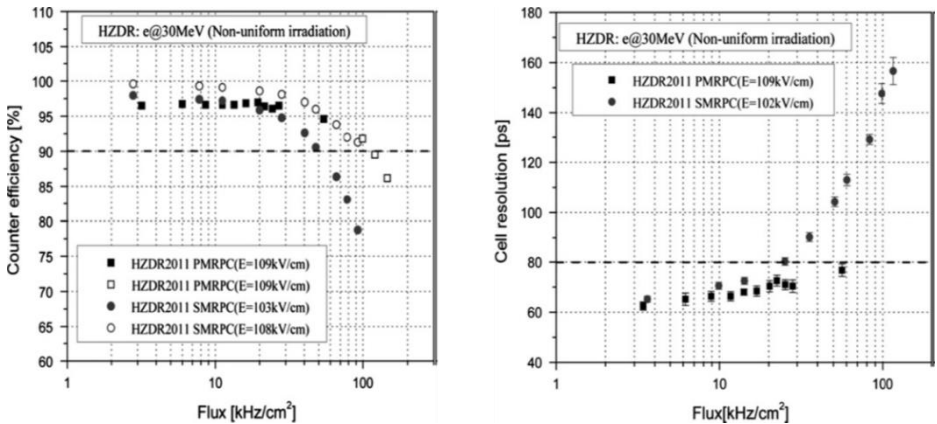

**Fig. 4.** Measured efficiencies and time resolutions for different runs as a function of the average particle flux determined with reference scintillators [8].

Based on the technique of the low-resistive glass, Tsinghua University has developed a double-ended readout strip MRPC, named MRPC3a in the TOF wall. It consists of two mirrored stacks of resistive plates, which fit into the three parallel readout PCBs. In each stack, the 0.25 mm nylon monofilaments spacers divide the five low-resistive glass plates into four homogeneous gas gaps. The top and bottom plates among these five glasses are covered with the colloidal graphite spray and used as the high voltage electrode. There are 32 strips on each readout PCB. Each of them is 270 mm long, 7 mm wide and the interval is 3 mm. Ground is placed onto the MRPC's electrode. Feed through is carefully designed to match the 100 Ω impedance of PADI electronics. This method aims to minimize the noise caused by reflection. The photo of the detector is shown in Fig.5.

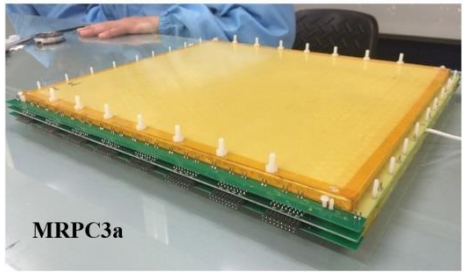

**Fig. 5.** Photo of the MRPC3a prototype designed by Tsinghua University.

The performances of MRPC3a are shown in Fig 5 [9]. It can be seen the efficiency stays at 97%. The cluster size always maintains a low value of 1.6. Assuming an equal performance with the reference counter we have a time resolution of the MRPC3a of about 50 ps.

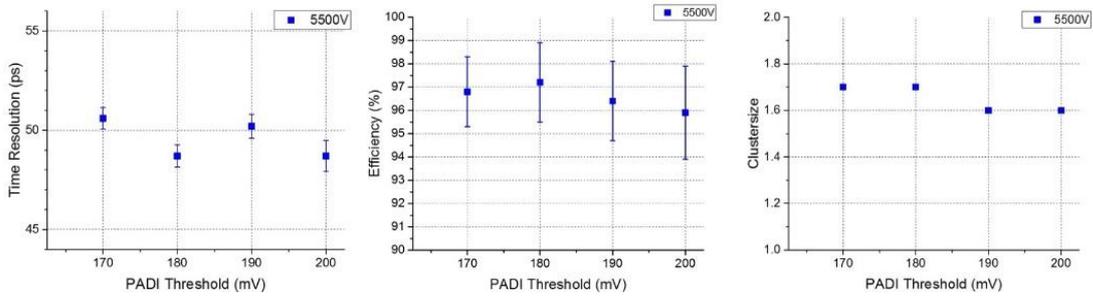

**Fig. 6.** Time resolution (around 50 ps), efficiency (97%) and clustersize (1.6 to 1.7) of MRPC3a under different FEE (PADI) electronics threshold [9].

The electronics is nearly the same as used in STAR-TOF, The ASIC differential amplifier is PADI [10] and ASIC TDC is GET4 [11]. The time jitter of PADI & GET4 is below 30ps. Now the CBM-TOF is in construction and the production of high rate MRPC and electronics goes smoothly. Fig.7 shows the mass production of high rate MRPC at Miyun workshop of Tsinghua University.

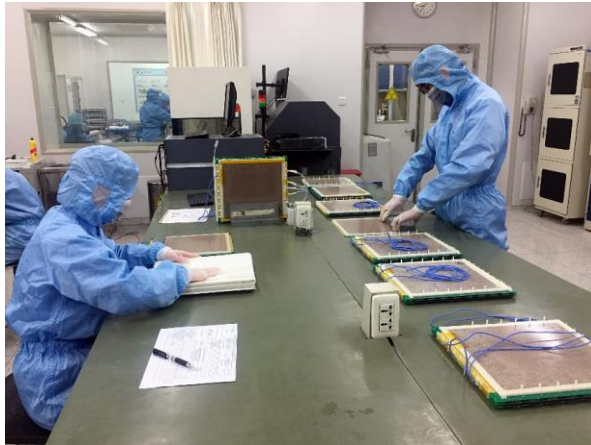
Fig.6 The mass production of high rate MRPC at Miyun workshop of Tsinghua University

## 4. The third generation TOF

The typical requirements for the third generation TOF are high time resolution (20ps) and high rate back ground (20kHz/cm$^2$). This technology first aimed for the particle identification of SoLID spectrometer at JLab. the spectrometer requires a TOF for Kaon/pion separation up to momentum of 7 GeV/c under high particle rate of 20 kHz/cm$^2$. As we know, the time resolution of TOF system is determined by the time jitter of MRPC and electronics, it can be shown in equation (1).

$$\sigma_{TOF} = \sqrt{\sigma_{MRPC} + \sigma_{electronics}} \tag{1}$$

If σ$_{TOF}$ is required to be less than 20ps, so the time jitter of MRPC and electronics should be smaller than 14ps. We have to develop very narrow gap MRPC. For the electronics, the time jitter of NINOs (PADI)+ HPTDC (GET4) is usually larger than 20ps. But the time jitter of fast amplifier+ pulse shape digitizer can meet the requirement. This technology route can be seen in Fig.7. The MRPC consists of 32 x0.104mm gaps, each stack consists of 8 gaps. The amplifier is developed in our institute and its bandwidth is around 350MHz. CAEN waveform digitizer DT5742 (based on DRS4-V5 Chip) is used to get pulse shape.

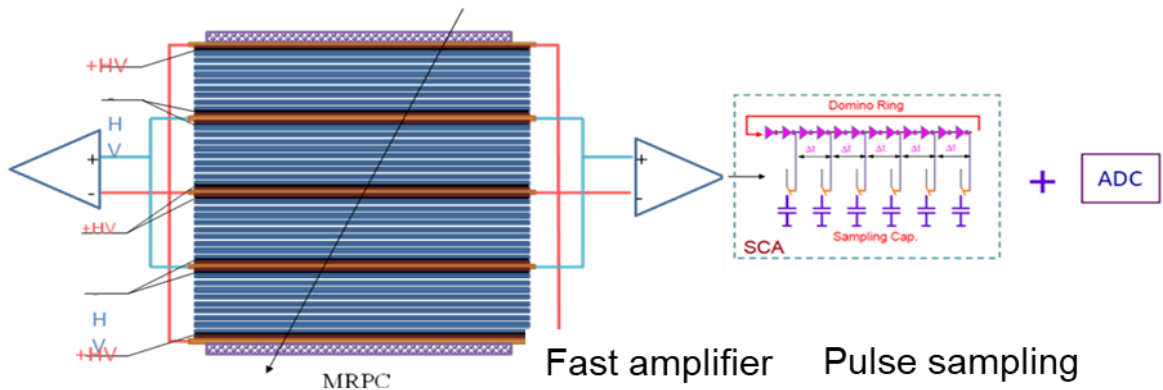
Fig.7 The high resolution of MRPC and readout chain.

Reconstructing the particle incident time (exactly the time point of first ionization) is always the most important job of the MRPC detector. Fig.8 shows the pulse shape and the timing point. This can be obtained from the simulation. Finding the incident time from the waveform is obviously a non-linear problem.

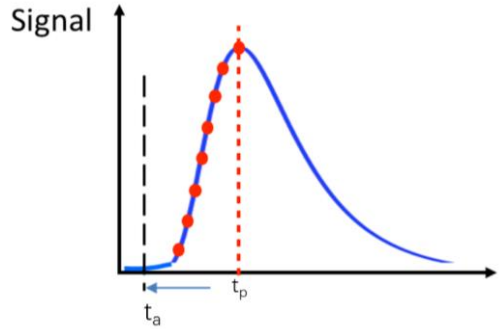

Fig.8 The exact particle incident time can be obtaned through the leading edge

In recent years, the artificial neural network(NN) has efficiently solved many non-linear problems not only in the field of the computer science, but also in high energy physics[12]. For the 3rd generation ToF system, we propose to use a fully-connected neural network to find out the patterns from the signal waveform and calculate the particle arriving time.[13] The network takes several uniformly distributed points on the leading edge of the signal waveform as the input and outputs the length of the leading edge $t_l$. By substracting $t_l$ from the peaktime $t_p$, the estimated particle arriving time is $t_a = t_p - t_l$. In order to train the NN, we simulate the pulsed signal of the MRPC shown in Fig7 in details[16]. 8 uniformly distributed points from the leading edge are extracted and fed to the neural network. Networks of different structures are trained and validated with the training(>120,000) and validation(>50,000) datasets.

Two identical MRPCs was constructed in the lab and tested with the cosmic rays. The sampling rate of the waveform digitizer is 5 GHz, so about 8 points along the leading edge are obtained. In the experiment, MRPCs are read out from both sides, and therefore 4 main waveforms(1L,1R,2L,2R) will be obtained for every single event. The incident of these 4 waveforms is estimated by the neural network model separately and for every event, $t_a$ is calculated. The estimated time for each MRPC is defined to be the average of the left and right. Fig.9 shows the time difference of two MRPCs and the time resolution is $38/\sqrt{2} = 27$ ps. If the time is calculated by the ToT method, a fixed threshold is set the waveform and $t_c$ and $t_{tot}$ is obtained.  After the slewing correction, the best result we achieve is only 48 ps. This result effectively proves that the neural network is a promising analysis algorithm to reconstruct the MRPC time.

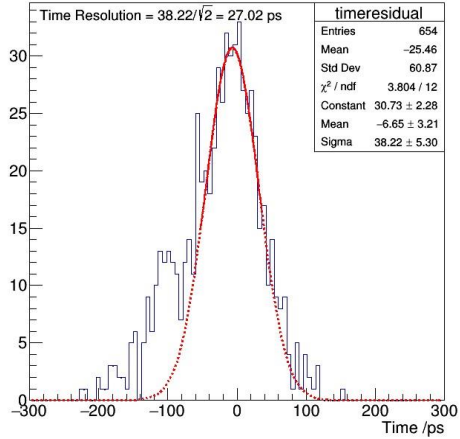

Fig.9 Time resolution of the test on experiment data

## 5. Conclusion

Three generation of MRPC TOF system were described. The main technology of them is shown in table 2. It can be seen with the increase of accelerator luminosity and PID precision, the requirement for TOF system becomes more stringent. High rate and high time resolution system is required for future physics experiments. It is necessary to develop new material and new structure MRPC detector. New kind electronics and analysis methods are also required to study.

Table 2. Main technology of three generation of MRPC TOF system

|  | Requirement | MRPC | Typical electronics | Analysis method | Typical experiment |
| --- | --- | --- | --- | --- | --- |
| 1st TOF | Time resolution<100ps Rate<100Hz/cm$^2$ | Gas gap:200-300 μm Electrode: float glass | NINOs +HPTDC | TOT slewing correction | RHIC-STAR LHC-ALICE |
| 2st TOF | Time resolution<100ps Rate~30kHz/cm$^2$ | Gas gap:200-300 μm Electrode: low resistive glass | PADI +GET4 | TOT slewing correction | FAIR-CBM |
| 3st TOF | Time resolution<20ps Rate~20kHz/cm$^2$ | Gas gap:100-150 μm Electrode: low resistive glass | Fast amplifier + pulse shape sampling | TOT slewing correction Deep learning technology | JLab-SoLID |

**Acknowledgments**

The work is supported by National Natural Science Foundation of China under Grant No.11420101004, 11461141011, 11275108, 11735009. This work is also supported by the Ministry of Science and Technology under Grant No. 2015CB856905, 2016 YFA0400100.